\documentclass[12pt] {article}
\pdfoutput=1
\usepackage{jheppreprint}
\usepackage{tikz}
\title{Realizability of the Lorentzian $(n,1)$-Simplex}
\author{Kyle Tate \textmd{and} Matt Visser}
\affiliation{School of Mathematics, Statistics, and Operations Research, \\
Victoria University of Wellington, PO Box 600, Wellington 6140, New Zealand}
\emailAdd{kyle.tate@msor.vuw.ac.nz}
\emailAdd{matt.visser@msor.vuw.ac.nz}
\abstract{In a previous article [JHEP 1111 (2011) 072; \texttt{arXiv:1108.4965}] we have developed a Lorentzian version of the Quantum Regge Calculus in which the significant differences between simplices in Lorentzian signature and Euclidean signature are crucial. In this article we extend a central result used in the previous article, regarding the realizability of Lorentzian triangles,  to arbitrary dimension. This technical step will be crucial for developing the Lorentzian model in the case of most physical interest: $3+1$ dimensions.

We first state (and derive in an appendix) the realizability conditions on the edge-lengths of a Lorentzian $n$-simplex in total dimension $n=d+1$, where $d$ is the number of space-like dimensions. We  then show that in any dimension there is a certain type of simplex which has all of its time-like edge lengths completely unconstrained by any sort of triangle inequality. This result is the $d+1$ dimensional analogue of the $1+1$ dimensional case of the Lorentzian triangle. 

\bigskip
\noindent
26 October 2011; 16 December 2011; \LaTeX-ed \today
}
\begin{document}
\maketitle
\newpage
\section{Introduction} \label{sec: intro}
In the formulation of Lorentzian Fixed Triangulations (LFT) which we developed in reference \cite{Tate:2011ct}, an important geometrical fact about a certain type of Lorentzian 2-simplex (Lorentzian triangle) in  $1+1$ dimensions is exploited: All three of its edge-lengths, one space-like and two time-like, are completely unconstrained by any form of the triangle inequalities. This is in stark contrast to the situation in Euclidean signature. In particular, the absence of any triangle inequality constraints allowed for the explicit demonstration, essentially dependent on the difference in configuration spaces, that the Euclidean and Lorentzian Quantum Regge Calculus are not simply related theories --- that is, they are not related by a naive ``Wick Rotation''. (See references~\cite{Regge:1961px, Williams:1996jb, Hamber:2011cn} for additional background.) 

In the case of quantum field theories defined on flat Minkowski space there are rigorous mathematical theorems relating the Euclidean-signature Osterwalder--Schrader axioms to the equivalent Lorentzian-signature Wightman axioms, so in that situation the choice of whether to work in Euclidean signature or Lorentzian signature can be based on technical convenience.  For quantum gravity no such formal axiom structure exists, and there is no equivalent theorem of this nature.  That is, there is no physics reason to expect candidate models for Euclidean quantum gravity and Lorentzian quantum gravity to be the same theory --- and in fact the difference in classical configuration spaces indicates that in (at least in 1+1 dimensions) they cannot be the same theory.

In this current article, we will demonstrate that in any dimension $n=d+1$ there is a Lorentzian $n$-simplex which has all of its time-like edge lengths unconstrained. This will be important for formulating LFT in $n$ dimensions, as it implies that the non-trivial distinction between the Lorentzian and Euclidean theories holds in any dimension. We also discuss the origin of constraints on the time-like edges of the remaining simplices which are not of the specific type mentioned above. 

\section{Realizability of generic simplices} \label{sec: volrea}

An $n$-simplex (in either Euclidean or Lorentzian signature), which we will denote by the symbol $<012 \cdots n>$, consists of $n+1$ vertices connected by $\frac12 n (n+1)$ edges. Given a set of $\frac12 n (n+1)$ squared edge lengths $\{ D^2_{ij} \,|\, i,j = 0,\cdots,n \}$, a simplex is called ``realizable'' if one can find a linear imbedding of the $n+1$ vertices into flat space/spacetime such that the squared distance from vertex $i$ to vertex $j$ is given by the specified $D^2_{ij}$. 
\begin{itemize}
\item 
For a Euclidean $n$-simplex this corresponds to finding a linear imbedding in the $\mathbb{R}^{n}$ Euclidean space, where the distance is given by the standard Euclidean inner product ($D^2_{ij} = (D^2_E)_{ij}$). 
\item
For a Lorentzian $n$-simplex this corresponds to finding a linear imbedding in the $\mathbb{R}^{d,1}$ Minkowski space with $n=d+1$, where the distance is given by the Minkowski inner product ($D^2_{ij} = (D^2_L)_{ij}$). 
\end{itemize}
For a Euclidean $n$-simplex a necessary and sufficient condition for it to be realizable in this sense is for its Cayley--Menger  determinant (and appropriate minors) to have the appropriate (dimension dependent) sign. This is equivalent to the statement that the square of the volume be positive~\cite{Blumenthal:1953aa, Sommerville, Cayley, Menger}, both for the simplex itself and for an appropriate selection of its sub-simplices. 

The so-called Cayley--Menger determinant is the determinant  of the $(n+2) \times (n+2)$ matrix $E_{ab}^2$ which has entries $E_{ij}^2 = \langle \vec{v}_{i} - \vec{v}_{j},\vec{v}_{i} - \vec{v}_{j} \rangle = (D_E^2)_{ij}$ for $i,j = 0,\cdots,n$, and which is augmented by an additional row and column with entries given by $E_{n+1,n+1}=0$ and $E_{i,n+1}=E_{n+1,j}=1$. That is
\begin{equation}
E^2_{ab} = \left[\begin{array}{c|c} (D_E^2)_{ij}  & 1_i \\ \hline 1_j & 0\end{array}   \right].
\end{equation}
We will call this matrix the Cayley--Menger matrix. The proportionality factor occurring in the volume formula is combinatorial in nature, and depends on the dimension $n$. The volume of the simplex is given by: 
\begin{equation}
\label{eq: EucCMdet}
\left(V_n\right)_E^2 = \frac{(-1)^{n+1}}{2^n (n!)^2} |E_{ab}^2|.
\end{equation}
The distance matrix $(D_E)^2_{ij}$ represents a realizable simplex only if 
\begin{equation}
\hbox{sign}\left(  |E_{ab}^2| \right) = (-1)^{n+1},
\end{equation}
so that the volume is real.

As stated in~\cite{Tate:2011ct} there is an analogous Cayley--Menger determinant and volume formula for the Lorentzian $n$-simplex, with Euclidean distances replaced by Lorentzian distances in the Lorentzian Cayley--Menger matrix $L^2_{ab}$. That is:
\begin{equation}
L^2_{ab} = \left[\begin{array}{c|c} (D_L^2)_{i,j}  & 1_i \\ \hline 1_j & 0\end{array}   \right].
\end{equation}
The corresponding Lorentzian Cayley--Menger determinant formula for the volume is 
\begin{equation}
\label{eq: LorCMdet}
\left(V_n\right)_L^2 = \frac{(-1)^{n}}{2^n (n!)^2} |L_{ab}^2|.
\end{equation}
The squared volume must again be positive for the simplex to be realizable in $\mathbb{R}^{d,1}$. Deriving this result is straightforward, and is very similar to the usual derivation in Euclidean space.  We will therefore relegate the details to appendix A. We should point out that the derivation presented in that appendix is ``direct'' --- the derivation does not mention the term ``Wick rotation'', nor does it need any such concept. In principle equation \eqref{eq: LorCMdet} describes a function of the ${1\over2}n(n+1)$ edge lengths whose positivity tells one everything that one needs to know about the realizability of Lorentzian $n$-simplices. 

However, its compact form obscures the main result of this article: There is a special class of Lorentzian simplices for which, (once the spacelike edges have been chosen in an appropriate manner), the function is always positive for any arbitrary choice of the time-like edge lengths. That is: those edge lengths which have $(D^2_L)_{ij}<0$ can be chosen arbitrarily.

\section{Realizability of the Lorentzian $(n,1)$ simplex} \label{sec: n1-simplex}

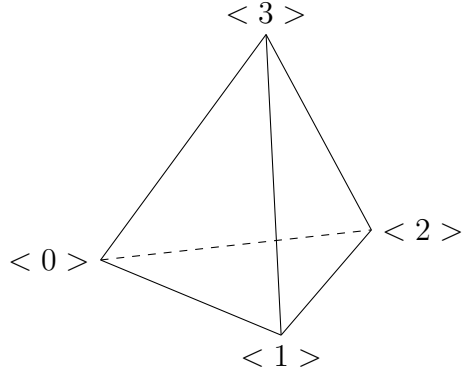
\begin{figure}[!htb]
\centering
\begin{tikzpicture}[scale=4.0]
\draw (0.2,0.2) node[anchor=north] {$<1>$} -- (-0.4,0.45) node[anchor = east] {$<0>$};
\draw (0.2,0.2) -- (0.5,0.55) node [anchor = west] {$<2>$};
\draw (0.2,0.2) -- (0.15,1.2) node [anchor = south] {$<3>$};
\draw (-0.4,0.45) -- (0.15,1.2);
\draw (0.5,0.55) -- (0.15,1.2);
\draw[dashed]  (-0.4,0.45)--(0.5,0.55);
\end{tikzpicture}
\caption{The (3,1) tetrahedron in 2+1 dimensions. The $<012>$ 2-simplex is a Euclidean triangle. The $<30>$, $<31>$, and $<32>$ 1-simplices are time-like edges whose lengths are unconstrained by any triangle inequalities.}
\label{F1}
\end{figure}

In both LFT \cite{Tate:2011ct} and CDT \cite{Ambjorn:1998xu, Loll:2000my, Ambjorn:2001cv, Ambjorn:2011kp, Ambjorn:2011cg}, the class of Lorentzian simplices one considers are the $(l,k)$ simplices (where $l+k=n+1$). These are simplices which have a Euclidean $(l-1)$-simplex in one space-like hypersurface, connected in ``time'' to a Euclidean $(k-1)$-simplex in the subsequent space-like hypersurface. Note that there are then
\begin{equation}
{l(l-1)\over2} +{k(k-1)\over2} =  {(n+1)(n-2k) + 2k^2\over 2}
\end{equation}
space-like edges and 
\begin{equation}
kl= k(n+1-k)
\end{equation}
time-like edges in the simplex. The simplex which in $n$ dimensions has all of its $n$ time-like edge lengths completely unconstrained by triangle inequalities is the $(n,1)$ simplex. (See for example figure~\ref{F1}.) 

Specifically, the $(n,1)$ simplex contains a Euclidean $(n-1)$-simplex sitting in an $n-1=d$ dimensional space-like hypersurface, which is connected by $n$ time-like edge lengths to a single vertex in the subsequent $n-1$ dimensional space-like hypersurface. We assume that the Euclidean $(n-1)$-simplex $<01\cdots(n-1)>$ is realizable, that is we can find coordinates $x_i^{\mu}$ for the $i=0,1,\cdots,n-1$ vertices with $x^{\mu} = (x^1,x^2,\cdots,x^n)$ where in this article we will denote the time coordinate by $t=x^n$. Without loss of generality we can choose our coordinate system such that the vertex $<0>$ is at the origin, then we can define the vectors $\vec{v}_1, \vec{v}_2, \cdots, \vec{v}_{n-1}$ to be the edges $<01>$,  $<02>$, $\cdots$, $<0(n-1)>$. Since these vectors span a space-like hypersurface, we can choose our coordinate system so that $x_i^{n} = t_i = 0$. In order for the $(n,1)$ simplex to be realizable, with time-like edge lengths given by
\begin{equation}
T_i^2 = \langle\vec{v}_n-\vec{v}_i,\vec{v}_n - \vec{v}_i\rangle\;<0,
\end{equation}
where $\vec{v}_n$ is the vector from the origin to the vertex $<n>$, we must be able to find coordinates $x_n^{\mu}$ such that the following nonlinear equations hold:
\begin{eqnarray}
\eta_{\mu \nu} (x_n-x_0)^{\mu}\;(x_n-x_0)^{\nu} &=& -T_0^2, \nonumber \\
\eta_{\mu \nu} (x_n-x_1)^{\mu}\;(x_n-x_1)^{\nu} &=& -T_1^2, \nonumber \\
\eta_{\mu \nu} (x_n-x_2)^{\mu}\;(x_n-x_2)^{\nu} &=& -T_2^2, \nonumber \\
 &\vdots& \nonumber \\
 \eta_{\mu \nu} (x_n-x_{n-1})^{\mu}\;(x_n-x_{n-1})^{\nu} &=& -T_{n-1}^2.
  \label{eq: Tsystemfirst}
\end{eqnarray}
We have chosen a coordinate system where $x_0^{\mu} = 0$, and we have boosted in the space-like hyper-surface so that $\forall i$ we have $x_i^{n} = 0$. We can make use of the rotational freedom in the space-like hypersurface to choose $x_i^{j} = 0$ for $i<j $, $i,j = 0,\cdots,n-1$. That is:
\begin{eqnarray}
x_1 &=& (x_1^1,0,0,\cdots,0), \nonumber\\
x_2 &=& (x_2^1,x_2^2,0,\cdots,0), \nonumber\\
x_3 &=& (x_3^1,x_3^2,x_3^3,\cdots,0), \nonumber\\
 &\vdots&\nonumber \\
x_{n-1} &=& (x_{n-1}^1,x_{n-1}^2,x_{n-1}^3,\cdots,x_{n-1}^{n-1},0).
\end{eqnarray}
In contrast the $n^{th}$ vector
\begin{equation}
x_{n} = (x_{n}^1,x_{n}^2,x_{n}^3,\cdots,x_{n}^{n-1},x_n^n)
\end{equation}
has no \emph{a priori} zeros. Note that the components $x_i^j$ form a lower-triangular matrix so the $n\times n$ determinant $|x_i^j|$ is simply $\prod_i x_i^i$. 
With these coordinate choices, the equations \eqref{eq: Tsystemfirst} become:
\begin{align}
&(x_n^1)^2 + (x_n^2)^2 + (x_n^3)^2 \cdots +(x_n^{n-1})^2 - (x_n^n)^2 = -T_0^2, \nonumber \\
&(x_n^1-x_1^1)^2 + (x_n^2)^2 + (x_n^3)^2 \cdots +(x_n^{n-1})^2 - (x_n^n)^2 = -T_1^2, \nonumber \\
&(x_n^1-x_2^1)^2 + (x_n^2-x_2^2)^2 + (x_n^3)^2 \cdots +(x_n^{n-1})^2 - (x_n^n)^2= -T_2^2, \label{eq: Tsystem} \\
 &\vdots \nonumber \\
&(x_n^1-x_{n-1}^1)^2 + (x_n^2-x_{n-1}^2)^2 + (x_n^3-x_{n-1}^3)^2 \cdots +(x_n^{n-1}-x_{n-1}^{n-1})^2 - (x_n^n)^2= -T_{n-1}^2. \nonumber
\end{align}
As long as the Euclidean sub-simplex is non-degenerate, which by the above requires that $\forall i\in\{1,\dots,n-1\}$ we have $x_i^i\neq 0$, this system of equations can be inverted recursively:
\begin{align}
& x_n^1 = \frac{ (x_1^1)^2 +T_1^2 - T_0^2 }{2 x_1^1}, \nonumber \\
& x_n^2 = \frac{ (x_2^2)^2 - 2x_n^1 x_2^1 + (x_2^1)^2 + T_2^2 - T_0^2}{2 x_2^2}, \nonumber \\
& x_n^3 = \frac{ (x_3^3)^2 - 2 x_n^2 x_3^2 + (x_3^2)^2 - 2 x_n^1 x_3^1 + (x_3^1)^2 + T_3^2 - T_0^2 }{2 x_3^3}, 
\label{eq: Tsystemsol} \\
& \vdots \nonumber \\
& x_n^n = t = \pm \sqrt{ (x_n^1)^2 + (x_n^2)^2 + (x_3^2)^2 + \cdots + (x_{n}^{n-1})^2 + T_0^2}. \nonumber
\end{align}
That the system of equations \eqref{eq: Tsystem} always has two distinct solutions (related by reflection through $t=0$) is best understood by realizing that it is equivalent to the fact that the $n$ hyperboloids defined in equations \eqref{eq: Tsystem} have two intersections, one in the joint intersection of the future light cones of the vertices $<0>,\cdots, <n-1>$, and one in the joint intersection of the past light cones. 

Thus we can state the main result of this article: \emph{In any dimension $n=d+1$,  $d>0$, there exists a Lorentzian $n$-simplex denoted $(n,1)$ which, provided its Euclidean sub-simplex is realizable, has all its $n$ time-like edge lengths completely unconstrained.} 

The result can also be understood by directly examining the Cayley--Menger determinant \eqref{eq: LorCMdet}. This is because, for the $(n,1)$ simplex all of the time-like edge lengths are only in one column (row). Suitably permuting indices one has
\begin{equation}
\label{eq: CMdetn1}
\left(V_{(n,1)}\right)^2 = \frac{(-1)^{n}}{2^n (n!)^2} \det(L^2) = \frac{(-1)^{n}}{2^n (n!)^2} 
\left| \; \begin{array}{c|c} E^2 & \vphantom{\big|} \,\vec{T} \\ \hline \vphantom{\Big|} \vec{T}^T & 0 \end{array} \; \right|.
\end{equation}
Here $E^2$ is the reduced Cayley--Menger matrix for the $d= (n-1)$ dimensional Euclidean sub-simplex, and 
\begin{equation}
\label{eq: vecT}
\vec{T} = \left( 1, -T_0^2 , -T_1^2, \cdots , -T_{n-1}^2 \right)^T.
\end{equation}
Let $U$ be the $(n+1) \times (n+1)$ identity matrix, augmented with an extra column and row, $U_{n+1,i} = \vec{\alpha}^T_i$, $U_{i,n+1} = 0$, and $U_{n+1,n+1} = \beta$, where the $\vec{\alpha}$ and $\beta$ are at this stage unspecified numbers:
\begin{equation}
\label{eq: Omat}
U = \left( \begin{array}{c|c} I & 0 \\ \hline \vphantom{\Big|} \vec{\alpha}^T & \beta \end{array} \right).
\end{equation}
Examine the matrix equation:
\begin{equation}
\label{eq: mateq}
\bar{L}^2  = U L^2 U^{T},
\end{equation}
where $\bar{L}^2$ is the Cayley--Menger matrix of the Euclidean $(n-1)$-simplex, augmented with an extra row and column, $\bar{L}^2_{n+1,n+1} = 1$, $\bar{L}^2_{n+1,i} = \bar{L}^2_{i,n+1} =0$:
\begin{equation}
\label{eq: barLdef}
\bar{L}^2 = \left( \begin{array}{c|c} E^2 & 0 \\ \hline 0 & 1 \end{array} \right).
\end{equation}
Solving equation \eqref{eq: mateq} is identical to solving the vector equation
\begin{equation}
\label{eq: veceq}
E^2 \vec{\alpha} = -\beta \vec{T}\,, 
\end{equation}
and the scalar equation
\begin{equation}
\label{eq: scaleq}
\beta \vec{\alpha}^T \vec{T} = 1.
\end{equation}
Since we are assuming that the Euclidean sub-simplex is realizable, (and in particular non-degenerate), $E^2$ is invertible and there is a unique solution to equation \eqref{eq: veceq} for $\vec{\alpha}$ in terms of $\beta$ and the $T_i^2$:
\begin{equation}
\label{eq: vecsol}
\vec{\alpha} = - \beta (E^2)^{-1} \vec{T}.
\end{equation}
Assuming that $\vec{T}^T (E^2)^{-1} \vec{T} \ne 0$, (we shall soon see that $\vec{T}^T (E^2)^{-1} \vec{T} = -2t^2$),  the scalar equation has solution
\begin{equation}
\label{eq: scalsol}
\beta^2 = - \frac{1}{\vec{T}^T (E^2)^{-1} \vec{T}}\,,
\end{equation}
where we have used the fact that $E^2 = (E^2)^T$. Note that $\det(U) = \beta$, and provided $\beta \ne 0$, then $U$ is invertible. Suppose this is the case, then we can rewrite equation \eqref{eq: CMdetn1} as
\begin{eqnarray}
\left(V_{(n,1)} \right)^2 &=& \frac{(-1)^{n}}{2^n (n!)^2} \det\left[ (U^{-1})^T \bar{L}^2 U^{-1} \right] \nonumber \\
 &=& \frac{(-1)^{n}}{2^n (n!)^2} \left( \det[U^TU] \right)^{-1} \det[ \bar{L}^2 ] \nonumber \\
 &=& \frac{(-1)^{n}}{2^n (n!)^2} \left( \det[U^TU] \right)^{-1} \det[ E^2 ] \nonumber \\
 &=& \frac{1}{2}  \frac{1}{\beta^2} \frac{1}{n^2}  \frac{(-1)^{(n-1)+1}}{2^{n-1} ((n-1)!)^2} \det[E^2] \nonumber\\
 &=& \frac{1}{2 \, \beta^2 \, n^2} (V_E)_{n-1}^2,
\end{eqnarray}
where, as stated above, $E^2$ is the $(n+1,n+1)$ cofactor of $\bar{L}^2$, which is the Cayley--Menger matrix for the Euclidean $(n-1)$-simplex, and in the last step we have used equation \eqref{eq: EucCMdet}. Finally, using the fact that the volume of a simplex can always be written as
\begin{equation}
V_n = \frac{1}{n} \; V_{n-1} \;h,
\end{equation}
where $h$ is the perpendicular distance from the $(n-1)$-sub-simplex to the last vertex,  and in light of equation \eqref{eq: Tsystemsol} we have:
\begin{equation}
\left( V_{(n,1)} \right)^2 = \frac{t^2}{n^2} (V_E)_{n-1}^2.
\end{equation}
Thus we can identify
\begin{equation}
\label{eq: alphanp1}
\beta = \frac{1}{\sqrt{2}\; t}.
\end{equation}
So we explicitly see that the positivity of the $(n,1)$ Cayley--Menger determinant reduces to the realizability condition on the Euclidean sub-simplex, plus the existence of a nonzero solution for $\beta$, or equivalently $t$. This existence of this solution is guaranteed by \eqref{eq: Tsystemsol}. As mentioned above, this decomposition can only be done for the $(n,1)$ simplex; for the other $(l,k)$ simplices the time-like edge lengths get mixed up in columns (rows) with the space-like edge lengths.

%
\section{Constraints on the other Lorentzian $(l,k)$ simplices}
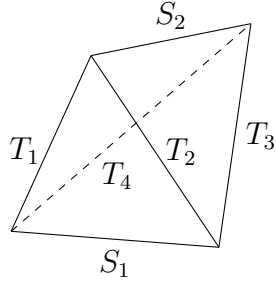
\begin{figure}[!htbp]
\centering
\begin{tikzpicture}[scale=4.25]
\draw (0,0)--(-0.65,0.05);
\draw (0,0)--(-0.4,0.6);
\draw (-0.65,0.05)--(-0.4,0.6);
\draw (0,0) -- (0.1,0.7);
\draw (-0.4,0.6)--(0.1,0.7);
\draw[dashed] (-0.65,0.05) -- (0.1,0.7);
\draw (-0.325,0.025) node [anchor = north] {$S_1$};
\draw (-0.2,0.3) node[anchor = west] {$T_2$};
\draw (-0.525,0.3025) node [anchor = east] {$T_1$};
\draw (0.05,0.350) node [anchor = west] {$T_3$};
\draw (-0.150,0.650) node [anchor = south] {$S_2$};
\draw (-0.4,0.3) node [anchor = north west ] {$T_4$};
\end{tikzpicture}
\caption{The (2,2) tetrahedron with edge lengths explicitly displayed. $S_1$ and $S_2$ are space-like, whereas $T_1$, $T_2$, $T_3$, and $T_4$ are time-like.}
\label{F2}
\end{figure}

To understand why the constraints on the $(n,1)$ simplex are straightforwardly simple, in contrast to those for the other $(l,k)$ simplices, it is instructive to consider the first dimension where this distinction arises: $n=2+1$. In dimension $2+1$ there are two simplices that need to be considered: The elementary $(3,1)$ simplex of figure~\ref{F1} and the more complex $(2,2)$ simplex of figure~\ref{F2}.

As stated above, the simplicity of the $(3,1)$ simplicial constraints can be seen to be a result of the fact that given a Euclidean triangle in a space-like hyper-surface the three hyperboloids, defined as the set of points for which the time-like distance from each vertex is fixed to a certain value, always have precisely two intersections: one in the forward light cone and one in the past lightcone.  For the $(2,2)$ simplex with edge lengths given by $S_1, S_2, T_1, T_2, T_3, T_4$ the fact that these edge legths may be non-realizable can be understood as follows: First, given $S_1$, $T_1$, $T_2$ we can always choose the coordinate system where the vertices are at $(x,y,t) =(-S_1/2,0,-S_1/2)$, $(S_1/2,0,-S_1/2)$, and $(X,0,\tau)$ respectively, where 
\begin{eqnarray}
\tau  &=& \frac{1}{2S_1} \left( \sqrt{(S^2 + [T_1 - T_2]^2)(S_1^2 + [T_1+T_2]^2)} - S_1^2 \right), \label{eq: tsol} \\
X &=& \frac{1}{2S_1} \left(T_2^2 - T_1^2\right). \label{eq: xsol}
\end{eqnarray}
Then, given some $S_2$, $T_3$, and $T_4$, we need to find coordinates $(x,y,t)$ such that we have a realizable $(2,2)$ simplex. For it to be realizable, the point $(x,y,t)$ must be inside the intersection of the light cones of the first two space-like vertices (call that region $\mathcal{T}_{12} = \mathcal{T}_1 \cap \mathcal{T}_2$), and outside of the lightcone of the third vertex (call that region $\mathcal{S}_3 = ( \mathcal{T}_3)^c =\mathbb{R}^{2,1} \setminus \mathcal{T}_3$). Thus we have:
\begin{eqnarray}
\mathcal{T}_{12} &=& \left\{ (t,x,y) \;\bigg|\; \left(|x| + \frac{S_1}{2}\right)^2 +y^2 - \left(t+ \frac{S_1}{2}\right)^2 \leq 0 \right\}; \\
\mathcal{S}_{3} &=& \left\{ (t,x,y) \;\big|\; (x-X)^2 + y^2 -(t-\tau)^2 >0 \right\}; \\
\mathcal{R} &=& \mathcal{T}_{12} \cap \mathcal{S}_3. \label{E:region}
\label{eq: region}
\end{eqnarray}
Thus we see that the constraints on the $(2,2)$ simplex have their origin in the question of whether or not there exists a point $(x,y,t)$ in the region (\ref{E:region}) such that the three remaining edges (one space-like and two time-like) have length $S_2$, $T_3$ and $T_4$.

Of course this is equivalent to the fact, pointed out in reference \cite{Tate:2011ct}, that if one considers the time-like edge lengths as vectors directed from one of the space-like hyper-surface to the other, then $\vec{T}_1 + \vec{T}_3 = \vec{T}_2 + \vec{T}_4$. This is the same as stating that given $\vec{T}_1$ and $\vec{T}_2$ the other two time-like edges must be such that $\vec{T}_1- \vec{T}_4$ and $\vec{T}_2 - \vec{T}_4$ are space-like, that is they must be inside the region \eqref{eq: region}. 

Turning to the physically interesting case of $3+1$ dimensions, there are only two distinct $(l,k)$ simplices: The $(4,1)$ simplex, which by the results of this article has no constraints on its time-like edge lengths, and the $(3,2)$ simplex. This $(3,2)$ simplex in 3+1 dimensions will be constrained in a similar fashion to the $(2,2)$ simplex in 2+1 dimensions. Note that the $(3,2)$ simplex contains one Euclidean triangle, nine Lorentzian triangles, two $(3,1)$ tetrahedra and three $(2,2)$ tetrahedra. Each of these sub-simplices must be realizable and hence, in addition to a standard Euclidean constraint, (that on the Euclidean triangle), there will also be constraints on the six time-like edge lengths arising from the $(2,2)$ tetrahedral sub-simplices. Note however that in general for a simplex to be realizable it is necessary but not sufficient that its sub-simplices be realizable. That is to say, one should expect that in addition to the constraints arising from the $(2,2)$ tetrahedron, there will now be further constraints arising from the $3+1$ dimensional Cayley--Menger determinant. Nevertheless, it is safe to say that for certain choices of time-like edges the Lorentzian $(3,2)$ simplex will not be realizable, which stands in contrast to the Lorentzian $(4,1)$ simplex.

\section{Conclusions} \label{sec: dihed}
In this article we have shown that in any dimension $n=d+1$,  where $d>0$, there exists a Lorentzian $n$-simplex denoted $(n,1)$ which, provided its Euclidean sub-simplex is realizable, has its $n$ time-like edge lengths completely unconstrained. This was claimed in reference \cite{Tate:2011ct} without explicit proof, and this article has served as verification of that claim. As a result we have shown that in all dimensions, the geometry of simplices in Lorentzian signature can be surprisingly and significantly different from their geometry in Euclidean signature.

 This observation is of particular relevance for simplicial models of quantum gravity in that it implies that the configuration space for path integrals over simplicial manifolds, that is the set of edge lengths for which every simplex in the manifold is realizable, depends on whether one's model is formulated in Euclidean or Lorentzian signature. This difference further strengthens the notion, first encountered in the formulation of CDTs and further developed in reference \cite{Tate:2011ct}, that simplicial approaches to quantum gravity should be developed directly in Lorentzian signature, rather than first in Euclidean signature and then trying to Wick rotate to relate the results to Lorentzian signature. 

\appendix
\section[Appendix: Lorentzian-signature Cayley--Menger determinant]
{Appendix:\\ Lorentzian-signature Cayley--Menger determinant} \label{sec: appendix}
Suppose we have a realizable $n$-simplex in $n=d+1$ Minkowski space $<012...n>$, with squared edge lengths $(D_L^2)_{ij}$. By realizability, we mean that given the $n+1$ edge lengths there exists a linear embedding of the simplex in $d+1$ space, that is we can find coordinates $x_i^{\mu}=(x_i^1,...,x_i^d,t_i)$ for each vertex $i=0,...,n$ (where we identify $t_i = x_i^n$) such that $(D^2_L)_{ij} = \langle x_i-x_j,x_i-x_j \rangle$. If this holds, and we place the vertex $0$ at the origin, then exactly as if this were a Euclidean simplex, the volume is given by
\begin{equation}
V_n = \frac{1}{n!} \det X_{ij},
\end{equation}
where $X_{ij} = x^{j}_i$. Let us now use this result to derive the Lorentzian-signature Cayley--Menger formula for Lorentzian volume. We have:
 \begin{equation}
V_n = \frac{1}{n!} \left| \begin{array}{ccc} 
x_1^1 & \cdots & x_1^n \\
\vdots&\ddots & \vdots \\
x_n^1 & \cdots & x_n^n \end{array} \right|.
\end{equation}
We can move away from referencing the origin by subtracting the coordinates of $x_0^{\mu}$:
 \begin{equation}
V_n = \frac{1}{n!} \left| \begin{array}{ccc} 
x_1^1-x_0^1 & \cdots & x_1^n-x_0^n \\
\vdots&\ddots & \vdots \\
x_n^1-x_0^1 & \cdots & x_n^n-x_0^n \end{array} \right|.
\end{equation}
We can add the following column and row without changing the volume, changing the determinant by $(-1)^{n+1}$:
 \begin{equation}
V_n = \frac{(-1)^{n+1}}{n!} \left| \begin{array}{cccc} 
1&x_1^1-x_0^1 & \cdots & x_1^n-x_0^n \\
\vdots&\vdots&\ddots & \vdots \\
1&x_n^1-x_0^1 & \cdots & x_n^n -x_0^n\\
1&0 &\cdots &0 \end{array} \right|.
\end{equation}
Adding $x_0^{i}$ times the first column to the $i^\text{th}$ column does not change the determinant:
 \begin{equation}
V_n = \frac{(-1)^{n+1}}{n!} \left| \begin{array}{cccc} 
1&x_1^1 & \cdots & x_1^n \\
\vdots&\vdots&\ddots & \vdots \\
1&x_n^1 & \cdots & x_n^n \\
1&x_0^1 &\cdots &x_0^n \end{array} \right|.
\label{E:A5}
\end{equation}
Multiply the last column by $-1$:
\begin{equation}
V_n = \frac{(-1)^{n}}{n!} \left| \begin{array}{cccc} 
1&x_1^1 & \cdots & -x_1^n \\
\vdots&\vdots&\ddots & \vdots \\
1&x_n^1 & \cdots & -x_n^n \\
1&x_0^1 &\cdots &-x_0^n \end{array} \right|,
\label{E:A6}
\end{equation}
then multiply each side of (\ref{E:A6}) by the determinant of the transpose of (\ref{E:A5}):
\begin{equation}
V_n^2 = \frac{-1}{(n!)^2} \left| \begin{array}{cccc} 
1&x_1^1 & \cdots &- x_1^n \\
\vdots&\vdots&\ddots & \vdots \\
1&x_n^1 & \cdots & -x_n^n \\
1&x_0^1 &\cdots &-x_0^n \end{array} \right| \;  \left| \begin{array}{cccc} 
1&1 & \cdots & 1\\
x_1^1&\cdots&x_n^1 &  x_0^1 \\
\vdots&\vdots & \ddots & \vdots \\
x_1^n&\cdots &x_n^n &x_0^n \end{array} \right|.
\end{equation}
Hence
\begin{equation}
V_n^2 = \frac{-1}{(n!)^2} \left| \begin{array}{cccc} 
1+ \langle x_1,x_1 \rangle&1+ \langle x_1,x_2 \rangle & \cdots & 1+ \langle x_1,x_0 \rangle \\
\vdots&\vdots&\ddots & \vdots \\
1+ \langle x_n,x_1 \rangle &1+ \langle x_n,x_2 \rangle & \cdots & 1+ \langle x_n,x_0 \rangle \\
1+ \langle x_0,x_1 \rangle &1+ \langle x_0,x_2 \rangle &\cdots &1+ \langle x_0,x_0 \rangle \end{array} \right|,
\end{equation}
where $\langle \cdot,\cdot\rangle$ is the Minkowski inner product. Now add the following rows and columns, noting that they do not change the determinant:
\begin{equation}
V_n^2 = \frac{-1}{(n!)^2} \left| \begin{array}{cccc} 
1&1 &\cdots&1\\
0&1+\langle x_1,x_1\rangle& \cdots & 1+\langle x_1,x_0\rangle \\
\vdots&\vdots&\ddots & \vdots \\
0&1+\langle x_n,x_1\rangle& \cdots & 1+\langle x_n,x_0\rangle \\
0&1+\langle x_0,x_1\rangle&\cdots &1+\langle x_0,x_0\rangle \end{array} \right|.
\end{equation}
Subtract the first row from every other row:
\begin{equation}
V_n^2 = \frac{-1}{(n!)^2} \left| \begin{array}{cccc} 
1&1 &\cdots&1\\
-1&\langle x_1,x_1\rangle& \cdots & \langle x_1,x_0\rangle \\
\vdots&\vdots&\ddots & \vdots \\
-1&\langle x_n,x_1\rangle& \cdots & \langle x_n,x_0\rangle \\
-1&\langle x_0,x_1\rangle&\cdots &\langle x_0,x_0\rangle \end{array} \right|.
\end{equation}
The cofactor $C_{11}$ of the top-left element  is zero since:
\begin{equation}
\left| \begin{array}{ccc} 
\langle x_1,x_1\rangle& \cdots & \langle x_1,x_0\rangle \\
\vdots&\ddots & \vdots \\
\langle x_n,x_1\rangle& \cdots & \langle x_n,x_0\rangle \\
\langle x_0,x_1\rangle&\cdots &\langle x_0,x_0\rangle \end{array} \right| 
= 
\left| \begin{array}{cccc} 
0&x_1^1 & \cdots & -x_1^n \\
\vdots&\vdots&\ddots & \vdots \\
0&x_n^1 & \cdots & -x_n^n \\
0&x_0^1 &\cdots & -x_0^n 
\end{array} \right| \;  
\left| \begin{array}{cccc} 
0&0 & \cdots & 0\\
x_1^1&\cdots&x_n^1 &  x_0^1 \\
\vdots&\vdots & \ddots & \vdots \\
x_1^n&\cdots &x_n^n &x_0^n \end{array} \right| = 0.
\end{equation}
Thus setting the top-left $11$ element to $0$, and then multiplying the first column by $-1$,  the only effect is that the volume determinant simply changes sign:
\begin{equation}
V_n^2 = \frac{1}{(n!)^2} \left| \begin{array}{cccc} 
0&1 &\cdots&1\\
1&\langle x_1,x_1\rangle& \cdots & \langle x_1,x_0\rangle \\
\vdots&\vdots&\ddots & \vdots \\
1&\langle x_n,x_1\rangle& \cdots & \langle x_n,x_0\rangle \\
1&\langle x_0,x_1\rangle&\cdots &\langle x_0,x_0\rangle \end{array} \right|.
\end{equation}
Multiply every column by $-2$ except the first (that is, multiply the last $n+1$ columns by $-2$), and then multiply the first row by $-1/2$:
\begin{equation}
V_n^2 = (-2)\; \frac{1}{(-2)^{n+1}}\frac{1}{(n!)^2} \left| \begin{array}{cccc} 
0&1 &\cdots&1\\
1&-2\langle x_1,x_1\rangle& \cdots & -2\langle x_1,x_0\rangle \\
\vdots&\vdots&\ddots & \vdots \\
1&-2\langle x_n,x_1\rangle& \cdots & -2\langle x_n,x_0\rangle \\
1&-2\langle x_0,x_1\rangle&\cdots &-2\langle x_0,x_0\rangle \end{array} \right|.
\end{equation}
Finally add $\langle x_i,x_i\rangle$ times the first column to the $(i+1)^{th}$ column, and $\langle x_i,x_i\rangle$ times the first row to the $(i+1)^{th}$ row, (when $i=n+1$ multiply instead by  $\langle x_0, x_0 \rangle$).
Then:
\begin{eqnarray}
V_n^2 &=& \frac{1}{(-2)^{n}(n!)^2} \times \\ && \left| \begin{array}{cccc} 
0&1 &\cdots&1\\
1&\langle x_1,x_1\rangle-2\langle x_1,x_1\rangle+\langle x_1,x_1\rangle& \cdots &\langle x_1,x_1\rangle -2\langle x_1,x_0\rangle+\langle x_0,x_0\rangle \\
\vdots&\vdots&\ddots & \vdots \\
1&\langle x_1,x_1\rangle-2\langle x_n,x_1\rangle+\langle x_n,x_n\rangle& \cdots &\langle x_n,x_n\rangle -2\langle x_n,x_0\rangle+\langle x_0,x_0\rangle \\
1&\langle x_1,x_1\rangle-2\langle x_0,x_1\rangle+\langle x_0,x_0\rangle&\cdots &\langle x_0,x_0\rangle-2\langle x_0,x_0\rangle +\langle x_0,x_0\rangle\end{array} \right| \nonumber \\
 &=&  \frac{1}{(-2)^{n}(n!)^2} \left| \begin{array}{cccc} 
0&1 &\cdots&1\\
1&0& \cdots &L_{10}^2\\
\vdots&\vdots&\ddots & \vdots \\
1&L_{n1}^2& \cdots &L_{n0}^2 \\
1&L_{01}^2&\cdots &0\end{array} \right|.
\end{eqnarray}
That is
\begin{equation}
V_n^2  = \frac{(-1)^{n}}{2^n (n!)^2} |L_{ij}^2|.
\end{equation}
This completes the derivation of the Lorentzian signature Cayley--Menger result.


\end{document}